\documentclass[]{interact}

\usepackage{epstopdf}
\usepackage{subfigure}
\usepackage{multirow}
\usepackage{float}
\usepackage[numbers,sort&compress]{natbib}
\usepackage{xcolor}
\usepackage{adjustbox}
\bibpunct[, ]{[}{]}{,}{n}{,}{,}

\theoremstyle{plain}

\theoremstyle{definition}

\theoremstyle{remark}

\begin{document}

\articletype{ARTICLE TEMPLATE}

\title{Generalizing the normality: a novel towards different estimation methods for skewed information}

\author{ Diego C Nascimento$^{\rm a}$, Pedro Luiz Ramos$^{\rm b}$, David Elal-Olivero$^{\rm a}$, Milton Cortes-Araya$^{\rm a}$, Francisco Louzada$^{\rm b}$ \\ $^{a}${Departamento de Matemática, Facultad de Ingeniería, Universidad de Atacama, Copiapó 1530000, Chile} \\ $^{b}${Institute of Mathematical Science and Computing, University of S\~ao Paulo, S\~ao Carlos, Brazil}}


\maketitle

\begin{abstract}
Normality is the most often mathematical supposition used in data modeling. Nonetheless, even based on the law of large numbers (LLN), normality is a strong presumption given that the presence of asymmetry and multi-modality in real-world problems is expected. Thus, a flexible modification in the Normal distribution proposed by Elal-Olivero \cite{elal2010alpha} adds a skewness parameter, called Alpha-skew Normal (ASN) distribution, enabling bimodality and fat-tail, if needed, although sometimes not trivial to estimate this third parameter (regardless of the location and scale). This work analyzed seven different statistical inferential methods towards the ASN distribution on synthetic data and historical data of water flux from 21 rivers (channels) in the Atacama region. Moreover, the contribution of this paper is related to the probability estimation surrounding the rivers' flux level in Copiapó city neighborhood, the most important economic city of the third Chilean region, and known to be located in one of the driest areas on Earth, besides the North and the South Pole.

\end{abstract}

\section{Introduction}
We live in the Big Data Era, in which a high volume and variety of data characterization are often noticeable in a data lake, nonetheless despite its amount of observation, symmetry, and smooth-tail are not always observed. These characteristics are natural since we all live in a complex world, with nonlinear relations and outliers, describing extreme values more recurrent than easy statistical tools take into account. This new age requires flexible models and different reasoning based on data information.

An often question crossing some traditional departments worldwide, "Are Statistics methods getting old fashion?". Sir David Cox \cite{cox2018big} explains that the focus is on the data relevance and quality, based on its coverage and representativeness, which gives confidence for the results, despite the amount of information (large volume) in the set which may hold some potentially biased estimates, with measurement errors. Efron \& Hastie \cite{efron2016computer} discuss the relation across computer-related and statistical inference as a mathematical logic system for guidance and correction, complemented by the large-scale prediction algorithms, suitable for this new century.

Therefore, complexity is intrinsic to massive data where high-dimension is often presented, and dynamic \cite{leonelli2020coherent, smith1987decision}. Nonetheless, all the information contained in the acquired data can be extracted through an estimation method, i.e., in maximum likelihood estimation  (MLE), and in a parametric version, it will be supported by a supposed distribution. Parametric approaches are easy to interpret patterns through parameters and enable association across variables, present a low computational cost, and be easier to implement in decision-making systems. 

In many cases, the standard MLE may not return desirable results. Other estimation methods that return accurate estimates have been considered, such as estimators based on the least square function \cite{swain1988least}, the product of spacing \cite{cheng1979maximum,ranneby1984maximum} or goodness-of-fit statistics \cite{luceno2006fitting}. There is no unique method that performs better for all models and it may depend on the selected parametric form \cite{ramos2019modeling, louzada2020exponential}.  Thus, using an efficient estimation method jointly with a flexible parametric model that covers many data patterns is demanded, which sometimes accommodates asymmetry and multi-modality that may be contained in the data That is considered.

This is the case of meteorological data, which shows significant changes as well as a complex dynamic \cite{bonnail2018trapping, du2019precipitation}. These field demands an extra attention, on data-driven models, thus needed to incorporate spatial-time dependence \cite{lopes2008spatial}, structural change \cite{mutti2020ndvi}, extreme value \cite{dutfoy2014multivariate}, but moreover in the parametric world a model supported by a probabilistic model (which deals with asymmetry and multi-modality \cite{ramoslouzada2016, rodrigues2018poisson}). Thus, this paper was motivated by the study case of the water flux of 21 rivers (channels) from the surroundings of Copiapó city, placed in the Atacama Desert, as one of the planet's driest areas. Moreover, we are persuaded to exemplify evidence towards the probability density associated with these events' empirical distribution through seven different statistical inference approaches.

This paper is divided into four parts. Section \ref{data} presents the motivation and details regarding the analyzed data. Section \ref{theorical} provides a background about the adopted methodology towards statistical inference elements. Then, Sections \ref{simul} and \ref{results} show the results related to the implemented methodology on synthetic data and the real-world data analysis. Finally, Section \ref{conclusion} discusses the conclusions based on the obtained results.

\section{The Data}\label{data}

The adopted data set is related to Fluviometric records (average monthly flows), from the Atacama Desert region (third region of Chile), in the Copiapó city neighborhood. The historical period of these data are from the past ten years from Jan, 2011 to Dec, 2020 associated with 21 rivers (or stream channel), obtained from the Chilean government web-site called \textit{Direccion General de Aguas} (\textit{Información Oficial Hidrometeorológica y de Calidad de Aguas en Línea}).

Historical events reveal the high periodicity of the low water flux of the region. However, cyclical events were also noticeable (such as two showers of rain, defrost of glaciers/snow in summer, amongst others), creating an expected multi-modality and large leptokurtosis.

\begin{figure}
    \centering
    \includegraphics[scale=0.35]{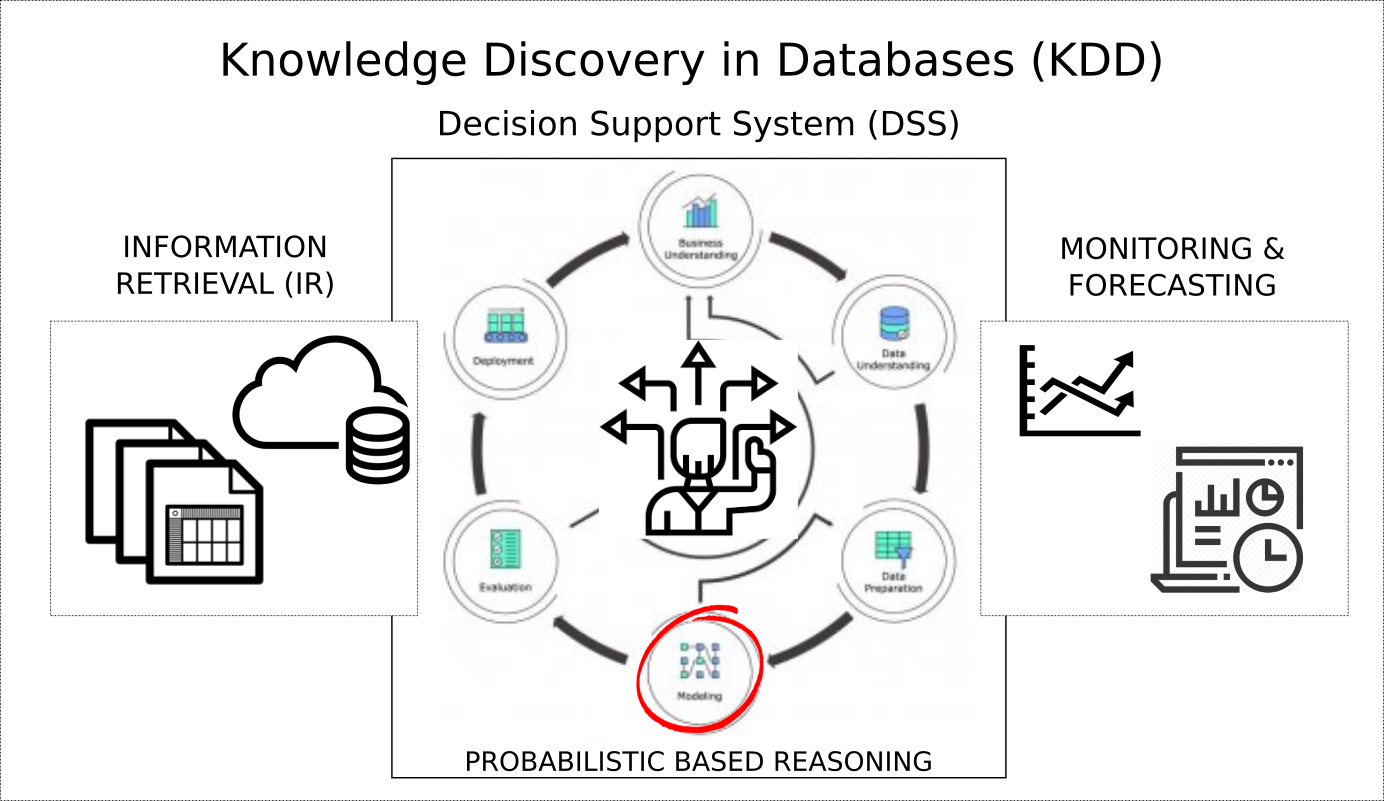}
    \caption{Visual summary of the probabilistic role in the Knowledge discovery in databases, as a cornerstone for quantification of uncertainty. Statistical inference procedures enable us to draw conclusions based on a sample, generalizing to the entire population.}
    \label{fig:DSS}
\end{figure}

Within the process information, a decision support system (DSS) sifts through and analyzes massive amounts of data, compiling comprehensive information that can be used to solve problems and in decision-making. Figure \ref{fig:DSS} summarizing chart flow of the knowledge discovery in databases (KDD), from the information retrieval (IR), decision support system (DSS), to monitoring \& forecasting.

Thus, the Atacama Desert watershed problem is one of a multi-dimensional study related to the circular economy to be analyzed. It is essential to mention that uncertainty is always presented globally (as a measurement error, sample bias, amongst others). Nonetheless, probabilistic reasoning allows one to generalize results through statistical inference procedures.

\section{Statistical Inference Elements} \label{theorical}

\subsection{Alpha-skew Normal (ASN) distribution}

Let X be a random variable following a Alpha-skew Normal (ASN) distribution then its probability density function (PDF) is given by
\begin{equation*}
f(x|\alpha)=\frac{(1-\alpha x)^2+1}{2+\alpha^2}\phi(x)    
\end{equation*}
where $x\in\mathbb{R}$, $\alpha\in\mathbb{R}$ and $\phi(\cdot)$ is the PDF of the standard normal distribution. 

The cumulative density function (CDF) is given by
\begin{equation*}
F(x|\alpha)=\Phi(x)+\alpha\left(\frac{2-\alpha x}{2+\alpha^2}\right)\phi(x)    
\end{equation*}

Then, wrapping the ASN density f(x$| \alpha$) with the parameters for location ($\mu$) and scale ($\sigma$), that is, the random variable $T$ is defined by $T=\mu +\sigma X$, for $\mu\in\mathbb{R}$ and $\sigma>0$, given by:
\begin{equation}\label{PDF}
f(t|\mu,\sigma,\alpha)=\frac{(1-\alpha (t-\mu)\sigma^{-1})^2+1}{(2+\alpha^2)\sigma}\phi\left(\frac{t-\mu}{\sigma}\right).
\end{equation}
with CDF related to equation (\ref{PDF}) as
\begin{equation}\label{CDF}
F(t|\mu,\sigma,\alpha)=\Phi\left(\frac{t-\mu}{\sigma}\right)+\alpha\left(\frac{2\sigma-\alpha (t-\mu)}{(2+\alpha^2)\sigma}\right)\phi\left(\frac{t-\mu}{\sigma}\right)   .
\end{equation}

\begin{figure}[tbh]
\centering
\includegraphics[scale=0.47]{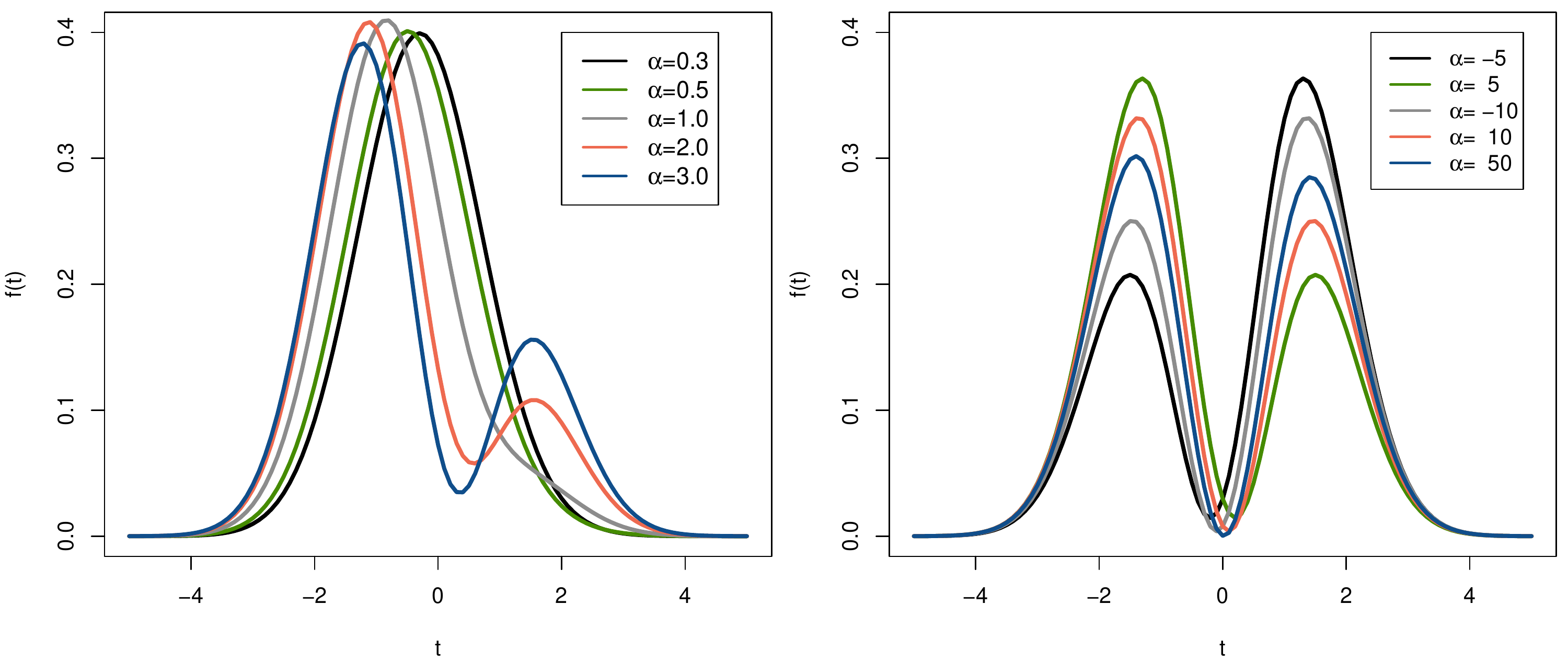}
\caption{PDF f(t) of the ASN distribution, where $t$ is a random variable, assuming $\mu=0$ (location), $\sigma=1$ (scale) and different values for $\alpha$.}
\label{fcv}
\end{figure}

Given its flexibility, the ASN distribution has been used in data modeling and adopted in different fields such as astronomy~\cite{tarnopolski2016analysis}, modeling wind speed~\cite{yang2018modeling}, and benchmark data~\cite{ara2019multivariate}. Figure \ref{fcv} presents different forms of the PDF of the ASN distribution, for instance, assuming $\mu=0$ (location), $\sigma=1$ (scale) and different values for $\alpha$, showing the presence of asymmetry and bimodality (incorporating different heights between the modalities);

\subsection{Different estimation methods for ASN distribution}

In this subsection, we will discuss seven different estimation methods (Maximum Likelihood Estimation, Ordinary and Weighted Least-Square Estimate, Method of Maximum Product of Spacings, Cramer-von Mises minimum distance estimators, Anderson-Darling and Right-tail Anderson-Darling estimators) for the parameters ($\mu, \sigma$, and $\alpha$) of the ASN distribution. Table \ref{TAB:methods} describe the methods used and the authors that proposed such inferential procedures. The comparison, which used the cited estimators, has been presented for other models \cite{deyramos2017, ramoslouzadab2018, teimouri}. 

\begin{table}[htbp]
\centering
\caption{Summarizing seven inferential estimation methods.}
\begin{adjustbox}{width=1\textwidth}
\small
\begin{tabular}{c|c|c}
\hline
\textbf{Estimation Method} & \textbf{Abbreviation} & \textbf{Created by} \\ \hline
Maximum Likelihood Estimation & MLE & Fisher \cite{fisher1922mathematical} \\ \hline
Ordinary Least-Square Estimate & LSQ & Swain et al. \cite{swain1988least}  \\ \hline
Weighted Least-Square Estimate & WLQ & Swain et al. \cite{swain1988least} \\ \hline
Maximum Product of Spacings & MPS & Cheng \& Amin \cite{cheng1979maximum} \\ \hline
Cramer-von Mises Estimators & CME & Macdonald \cite{macdonald1971estimation} \\ \hline
Anderson-Darling Estimator & ADE & Boos \cite{boos1982minimum}   \\ \hline
Right-tail Anderson-Darling Estimator & RADE & Luceno \cite{luceno2006fitting} \\ \hline
\end{tabular}
\end{adjustbox}
\label{TAB:methods}
\end{table}

Note that while Carl Friedrich Gauss introduced the LSQ in 1822 and it is one of the oldest estimation procedures, we have included the paper of Swain et al. \cite{swain1988least}
 The authors used such an approach for a class of non-normal models and became a standard reference when applied in different probability distributions. The additional details related to the estimation of the ASN distribution parameters are presented in the following subsections.

\subsubsection{Maximum Likelihood Estimation}

The Maximum Likelihood Estimation (MLE) is widely used in data analysis, where Fisher's derivation of the information inequality is seen at first for the analysis of variance, and later for estimate functions derived from Euler's Relation for homogeneous functions. Despite the fact that historical records of this technique have been widely exposed and defended by Ronald A. Fisher (maybe gained visibility because of the epic fights with Egon S. Pearson), its rationality dates back to the mid-1700s \cite{stigler2007epic}.

Let $t_{1}, t_{2},\cdots,t_{n}$ be the sample of the random sample of size $n$ from $F(\boldsymbol{t}| \mu,\sigma,\alpha)$. The maximum likelihood estimator $\hat{\mu}_{MLE}$, $\hat{\sigma}_{MLE}$ and $\hat{\alpha}_{MLE}$ can be obtained by maximizing, let's consider $z_i=\frac{(t_i-\mu)}{\sigma}$,
\begin{equation}\label{likehoodf}
L\left( \mu,\sigma,\alpha\right) = \frac{1}{(2+\alpha^2)^n\sigma^n}\prod_{i=1}^{n} \left((1-\alpha z_i)^2+1 \right)\phi\left(z_i\right),
\end{equation}

with respect to $\mu, \sigma $ and $\alpha$. The log-likelihood function of (\ref{likehoodf}) is given by
\begin{equation}\label{loglikehoodf}
\begin{aligned}
l\left( \mu,\sigma,\alpha\right) =& \sum_{i=1}^{n}\log \left((1-\alpha z_i)^2+1 \right)-n\log\left(2+\alpha^2 \right)-n\log(\sigma)+\sum_{i=1}^{n}\log\phi\left(z_i\right).
\end{aligned}
\end{equation}

From the expressions $\frac{\partial}{\partial \mu}l(\mu,\sigma,\alpha)=0$, $\frac{\partial}{\partial \sigma}l(\mu,\sigma,\alpha)=0$, $\frac{\partial}{\partial \alpha}l(\mu,\sigma,\alpha)=0$, the likelihood equations are
\begin{equation}\label{verogg21} 
\sum_{i=1}^{n}\frac{2\alpha(1-\alpha z_i)}{\left(1-\alpha z_i\right)^2+1} +\sum_{i=1}^{n} z_i=0.
\end{equation}

\begin{equation}\label{verogg22} 
\sum_{i=1}^{n}\frac{2\alpha z_i (1-\alpha z_i)}{(1-\alpha z_i)^2+1} +  \sum_{i=1}^{n} z_i ^2=0.
\end{equation}

\begin{equation}\label{verogg23} 
\sum_{i=1}^{n}\frac{2z_i(1-\alpha z_i)}{(1-\alpha z_i)^2 + 1} +\frac{2n\alpha }{1+\alpha^2}=0 .
\end{equation}
Numerical methods such as Newton-Rapshon are required to find the solution of the nonlinear system. Under mild conditions, the MLEs  are asymptotically normally distributed with a joint multivariate normal distribution given by
\begin{equation*}
(\hat{\mu}_{MLE},\hat{\sigma}_{MLE},\hat{\alpha}_{MLE})\sim N_3\left[(\mu,\sigma,\alpha),I^{-1}(\mu,\sigma,\alpha))\right] \mbox{ as } n \to \infty .
\end{equation*}
where $I(\mu,\sigma,\alpha)$ is the Fisher information matrix given in Elal-Olivero \cite{elal2010alpha}.

This methodology is often adopted given the MLE sufficiency, combined with consistency (which leads to the statistical efficiency), and asymptotic normality is guaranteed. Often cases, MLE are adopted under regularity conditions and assumptions like score functions linearly approximation. Nonetheless, some more constraints on the class of estimates (whenever the increase of parameters' numbers, unbounded likelihood functions, and the possibility of local improvement) should be verified and needed to be contours of the likelihood.

Next, we will present a series of Minimum Distance Estimations (easily applied to estimate consistently unknown parameters) and designed to reflect the proposed model reproducing the probabilistic structure of the real-world phenomenon under study \cite{wolfowitz1957minimum}. Minimum Distance Estimations provide consistent parameter estimates and competitive, especially when other methods did not succeed.

\subsubsection{Ordinary and Weighted Least-Square Estimate}
Let random sample of size $n$ present a sequence $t_{(1)}, t_{(2)},\cdots,t_{(n)}$ then been a series of order statistics in which $F(\boldsymbol{t}| \mu,\sigma,\alpha)$ is a monotonic function. The least square (LSQ) estimators $\hat{\mu}_{LSE}$, $\hat{\sigma}_{LSE}$ and $\hat{\alpha}_{LSE}$ for the ASN distribution can be obtained by minimizing the parameters $\mu, \sigma $ and $\alpha$, as 
\begin{equation*}
V\left( \mu,\sigma,\alpha\right) = \sum_{i=1}^{n}\left[ F\left( t_{(i)}\mid \mu,\sigma,\alpha \right) - \frac {i}{n+1} \right]^{2}.
\end{equation*}

Thus, the LSQ equations can be obtained through solving the non-linear equations
\begin{equation*}
\begin{aligned}
&\sum_{i=1}^{n}\left[ F\left( t_{(i)}\mid \mu,\sigma,\alpha \right) - \frac {i}{n+1}\right] \Delta_{j}\left( t_{(i)} \mid \mu,\sigma,\alpha \right) = 0, \quad  j=1,2,3 ,
\end{aligned}
\end{equation*}
where
\begin{equation}\label{delta1}
\begin{aligned}
\Delta_{1}\left( t_{(i)}\mid \mu,\sigma,\alpha \right) = &\frac{\partial}{\partial \mu} F\left( t_{(i)}\mid \mu,\sigma,\alpha \right) = \frac{\phi\left(z_i\right)}{(2+\alpha^2)\sigma}[2\alpha z_i - \alpha ^2 z_i^2 -2], \, \\
\Delta_{2}\left( t_{(i)}\mid \mu,\sigma,\alpha \right) = &\frac{\partial}{\partial \sigma} F\left( t_{(i)}\mid \mu,\sigma,\alpha \right)=\frac{z_i \phi\left(z_i\right)}{\sigma(2+\alpha^2)}[2\alpha z_i-\alpha^2 z_i^2 - 2], \\ 
\Delta_{3}\left( t_{(i)}\mid \mu,\sigma,\alpha \right) = &\frac{\partial}{\partial \alpha} F\left( t_{(i)}\mid \mu,\sigma,\alpha \right) = \frac{\phi\left(z_i\right)}{(2+\alpha^2)^2} [2-2\alpha^2 -4\alpha z_i].
\end{aligned}
\end{equation}%


Alternative solutions are obtained through numerical approximation, with high precision, for these $\Delta_{j}$ for $j=1,2,3$ partial derivatives.

Alternatively, the weighted least-squares (WLQ) estimates are proposed whenever efficient method is required under sets of small data, $\hat{\mu}_{WLSE}$, $\hat{\sigma}_{WLSE}$ and $\hat{\alpha}_{WLSE}$, can be obtained by minimized adopting the following equation,
\begin{equation*}
W\left( \mu,\sigma,\alpha \right) = \sum_{i=1}^{n}
\frac {\left( n+1\right)^{2}\left( n+2\right)}{i\left( n-i+1\right)}
\left[ F\left( t_{(i)}\mid \mu,\sigma,\alpha \right) - \frac {i}{n+1} \right]^{2}.
\end{equation*}
The solutions are deviated from the non-linear equations
\begin{equation*}
\sum_{i=1}^{n}\frac {\left( n+1\right)^{2}\left( n+2\right)}{i\left( n-i+1\right)}
\left[ F\left( t_{(i)}\mid \mu,\sigma,\alpha \right) -
\frac {i}{n+1} \right] \Delta_{j}\left( t_{(i)}\mid \mu,\sigma,\alpha\right) = 0, \quad j=1,2,3,
\end{equation*}
where $\Delta _{1}\left( \cdot \mid \mu,\sigma,\alpha \right)$, $\Delta _{2}\left( \cdot \mid \mu,\sigma,\alpha \right) $ and
$\Delta
_{3}\left( \cdot \mid \mu,\sigma,\alpha \right) $ are given in (\ref{delta1}). The WLQ estimation technique is particularly useful whenever one aims to weigh the observations proportional to the equivalence of the error variance for that observation, then overcoming the issue of non-constant variance.

\subsubsection{Method of Maximum Product of Spacings}

The maximum product of spacings (MPS) method is a powerful alternative to MLE for estimating unknown parameters of continuous univariate distributions, which aims to maximize the geometric mean of spacings in the data (differences between the values of the cumulative distribution function at neighborhood data points). Cheng \& Amin proposed this method \cite{cheng1979maximum, cheng1983estimating}, and also obtained independently by Ranneby \cite{ranneby1984maximum}, as a Kullback-Leibler information approximation measurement. Some desirable properties of the MPS methods such as asymptotic efficiency, invariance, and more importantly, consistency are held broadly (under general conditions) than for MLEs~\cite{cheng1983estimating}.

Let's represented the differences between the values of the cumulative distribution function on their neighborhood data points by the function $D_{i}(\mu,\sigma,\alpha)=F\left( t_{(i)}\mid\mu,\sigma,\alpha \right)
-F\left( t_{(i-1)}\mid \mu,\sigma,\alpha \right)$, for $i=\{1,2,\ldots ,n+1,\ldots \}$ as an uniform spacings of a random sample from the ASN distribution, defining by $F(t_{(0)}\mid \mu,\sigma,\alpha)=0$ and $F( t_{(n+1)} \mid \mu,\sigma,\alpha)=1$. The constraint of the
$\sum_{i=1}^{n+1} D_i (\mu,\sigma,\alpha) =1$ is held. Thus, the MPS estimates $\hat{\mu}_{MPS}$, $\hat{\sigma}_{MPS}$ and $\hat{\alpha}_{MPS}$ can be obtained by maximizing the geometric
mean of the spacings
\begin{equation}
G_{ASN}\left( \mu,\sigma,\alpha\right) =\left[ \prod\limits_{i=1}^{n+1}D_{i}( \mu,\sigma,\alpha)\right] ^{%
\frac{1}{n+1}} \label{G}
\end{equation}%
considering the maximization of this ($G_{ASN}$) function by adopting its logarithm as
\begin{equation}
H_{ASN}\left( \mu,\sigma,\alpha\right) =\frac{1}{n+1}\sum_{i=1}^{n+1}\log
D_{i} ( \mu,\sigma,\alpha).
\end{equation}

The estimates of the unknown parameters $\hat{\mu}_{MPS}$, $\hat{\sigma}_{MPS}$ and $\hat{\alpha}_{MPS}$ are obtained by solving the nonlinear equations
\begin{equation}
\frac{1}{n+1}%
\sum\limits_{i=1}^{n+1}\frac{1}{D_{i}(\mu,\sigma,\alpha))} \left[ \Delta_j
(t_{(i)} |  \mu,\sigma,\alpha) - \Delta_j (t_{(i-1)} |  \mu,\sigma,\alpha)
\right] =0, \quad j=1,2,3,
\end{equation}
where $\Delta _{1}\left( \cdot \mid \mu,\sigma,\alpha \right)$, $\Delta _{2}\left( \cdot \mid \mu,\sigma,\alpha \right) $ and
$\Delta_{3}\left( \cdot \mid \mu,\sigma,\alpha \right) $ are given respectively in (\ref{delta1}). 

It is important to mention that if $t_{(i+k)}=t_{(i+k-1)}=\ldots=t_{(i)}$ then $D_{i+k}(\mu,\sigma,\alpha)=D_{i+k-1}(\mu,\sigma,\alpha)=\ldots=D_{i}(\mu,\sigma,\alpha)=0$. Therefore, the MPS estimators are sensitive to closely spaced observations, especially ties. When the ties are due to multiple observations, $D_{i}(\mu,\sigma,\alpha)$ should be replaced by the corresponding likelihood $f(t_{(i)},\mu,\sigma,\alpha)$ since $t_{(i)}=t_{(i-1)}$. 

Under mild conditions for the ASN distribution, the MPS estimators are asymptotically normally distributed with a joint trivariate normal distribution given by
\begin{equation*}
(\hat{\mu}_{MPS},\hat{\sigma}_{MPS},\hat{\alpha}_{MPS})\sim N_3\left[(\mu,\sigma,\alpha),I^{-1}(\mu,\sigma,\alpha))\right] \mbox{ as } n \to \infty .
\end{equation*}

\subsubsection{The Cramer-von Mises minimum distance estimators}

Alternatively, an estimator that requires no assumption about the distributions' parametric form, the Cramer-von Mises estimator (CME), is based on the difference between the estimate of the cumulative distribution function and the empirical distribution
function \cite{cramer1928composition, von1928statistik}. These estimators operate based on the minimum distance across the "true" distribution (observed) and the "modeled" distribution (adjusted) through the maximum goodness-of-fit.

Macdonald \cite{macdonald1971estimation} showed that the bias of the estimator, from the CME, presents smaller distances alternatively to other minimum distance estimators. The Cramer-von Mises estimates $\hat{\mu}_{CME}$, $\hat{\sigma}_{CME}$ and $\hat{\alpha}_{CME}$ of the parameters $\mu$, $\sigma$ and $\alpha$ are obtained by minimizing through
\begin{equation}
C(\mu,\sigma,\alpha) =\frac{1}{12n}+\sum_{i=1}^{n}\left( F\left(
t_{(i)}\mid \mu,\sigma,\alpha\right) -{\frac{2i-1}{2n}}\right) ^{2}.
\end{equation}
Thus, these estimates are also obtained by solving the non-linear equations:
\begin{equation*}
\sum_{i=1}^{n}\left( F\left( t_{(i)}\mid \mu,\sigma,\alpha \right) -{\frac{2i-1}{2n}}\right) \Delta _{j}\left( t_{(i)}\mid \mu,\sigma,\alpha \right)  =0, \quad j=1,2,3,
\end{equation*}%
where $\Delta _{1}\left( \cdot \mid \mu,\sigma,\alpha \right)$, $\Delta _{2}\left( \cdot \mid \mu,\sigma,\alpha \right) $ and
$\Delta_{3}\left( \cdot \mid \mu,\sigma,\alpha \right) $ are given respectively in (\ref{delta1}).

\subsubsection{The Anderson-Darling and Right-tail Anderson-Darling estimators}

Another type of minimum distance estimator is based on Anderson-Darling statistics, often called Anderson-Darling estimator (ADE). This estimator is based on the minimum distance estimator obtained from sampling a sort data in ascending order of observed set (Y), then X = Sort(Y), and also combined with the permutation of $\{1, 2, \ldots, n\}$ which makes the X series sorted. Thus, this process is associated with the cumulative distribution function F$(\cdot)$ and the survival function S$(\cdot)=1-F(\cdot)$ for any PDF. In contrast, samples are drawn from a uniform distribution only if Y (and X) are samples from the PDF distribution.

The Anderson-Darling estimates $\widehat{\mu}_{ADE}, \widehat{\sigma}_{ADE}$ and $\widehat{\alpha}
_{ADE}$ of the parameters $\mu, \sigma$ and $\alpha$ are obtained
by minimizing, with respect to $\mu$, $\sigma$ and $\alpha$, the function
\begin{equation}
A(\mu,\sigma,\alpha) =-n-\frac{1}{n}\sum_{i=1}^{n}\left( 2i-1\right)
\left(\, \log F\left( t_{(i)}\mid \mu,\sigma,\alpha \right)+ \log S\left(
t_{(n+1-i)}\mid \mu,\sigma,\alpha\right) \, \right) .
\end{equation}
These estimates can also be obtained by solving the non-linear equations
\begin{equation*}
\sum_{i=1}^{n}\left( 2i-1\right) \left[ \frac{\Delta _{j}\left( t_{(i)}\mid
\mu,\sigma,\alpha \right) }{F\left( t_{(i)}\mid \mu,\sigma,\alpha \right) }-
\frac{\Delta_{j}\left( t_{(n+1-i)}\mid \mu,\sigma,\alpha \right) }{S\left( t_{(n+1-i)}\mid \mu,\sigma,\alpha \right) }\right] =0, \ \ j=1,2,3.
\end{equation*}

Alternatively, one can improve the ADE performance by taking into account the information held on the non-symmetrical differences between theoretical CDF and empirical CDF \cite{ye2017unilateral}. Thus, the Right-tail Anderson-Darling estimates (RADE) is an alternative though $\widehat{\mu }_{RADE}, \widehat{\sigma }_{RADE}$ and $\widehat{\alpha}_{RADE}$ of the parameters $\mu, \sigma$ and $\alpha$ are obtained by minimizing the function
\begin{equation}
R(\mu,\sigma,\alpha) =\frac{n}{2}-2\sum_{i=1}^{n}F\left( t_{i:n}\mid \mu,\sigma,\alpha \right) -\frac{1}{n}\sum_{i=1}^{n}\left( 2i-1\right) \log
S\left( t_{n+1-i:n}\mid \mu,\sigma,\alpha \right).
\end{equation}
These estimates can also be obtained by solving the non-linear
equations:
\begin{equation*}
- 2 \sum_{i=1}^{n} \Delta_{j}\left( t_{i:n}\mid \mu,\sigma,\alpha \right) +\frac{1}{n} \sum_{i=1}^{n}\left( 2i-1\right) \frac{\Delta_{j}\left( t_{_{n+1-i:n}}\mid \mu,\sigma,\alpha \right) }{S%
\left( t_{n+1-i:n}\mid \mu,\sigma,\alpha \right) } =0, \ \ j=1,2,3. 
\end{equation*}%
where $\Delta _{1}\left( \cdot \mid \mu,\sigma,\alpha \right)$, $\Delta _{2}\left( \cdot \mid \mu,\sigma,\alpha \right) $ and
$\Delta
_{3}\left( \cdot \mid \mu,\sigma,\alpha \right) $ are given respectively in (\ref{delta1}).\\

\section{Numerical Analysis}\label{simul}

In this section, we investigated the behavior of the ASN distribution based on artificial (synthetic) data, and its parameters modification conditioning on the estimation methodology. Thus, a Monte Carlo simulation was carried out, seven frequentist estimation methods were considered for the parameters, and comparing their efficiency. The following approach was adopted. The procedure is:
\begin{enumerate}
\item Generate $N$ samples of size $n$ given a set of parameters from the ${\rm ASN}(\mu, \sigma, \alpha)$ distribution;
\item For each generated set, based on the estimation methods (MLEs, LSQs, WLQs, MPSs, CMEs, ADEs and RTADEs, estimates of the parameters ($\mu$, $\sigma$ and $\alpha$) were calculated;
\item Then, considering $\boldsymbol{\hat\theta}=\left(\hat\mu,\hat\sigma,\hat\alpha\right)$ and $\boldsymbol{\theta}=\left(\mu,\sigma,\alpha\right)$ it was computed the Bias and Mean Squared Error (MSE) of $\boldsymbol{\hat\theta}$, which are given, respectively, by $\frac{1}{N}\sum_{k=1}^{N} \left(\hat{\theta}_{j}^{(k)}-\theta_{j}\right)$ and $\frac{1}{N}\sum_{k=1}^{N} \left(\hat{\theta}_{j}^{(k)}-\theta_{j}\right)^2$, for $j=\{ 1,2,3\}$ (each parameter). Whereas, $\hat\theta_{j}^{(k)}$ denotes the estimate of $\theta_j$ obtained from sample $k$, for $k=1,2,\cdots,N$.
\end{enumerate}

\begin{figure}[!t]
\centering
\includegraphics[scale=0.6]{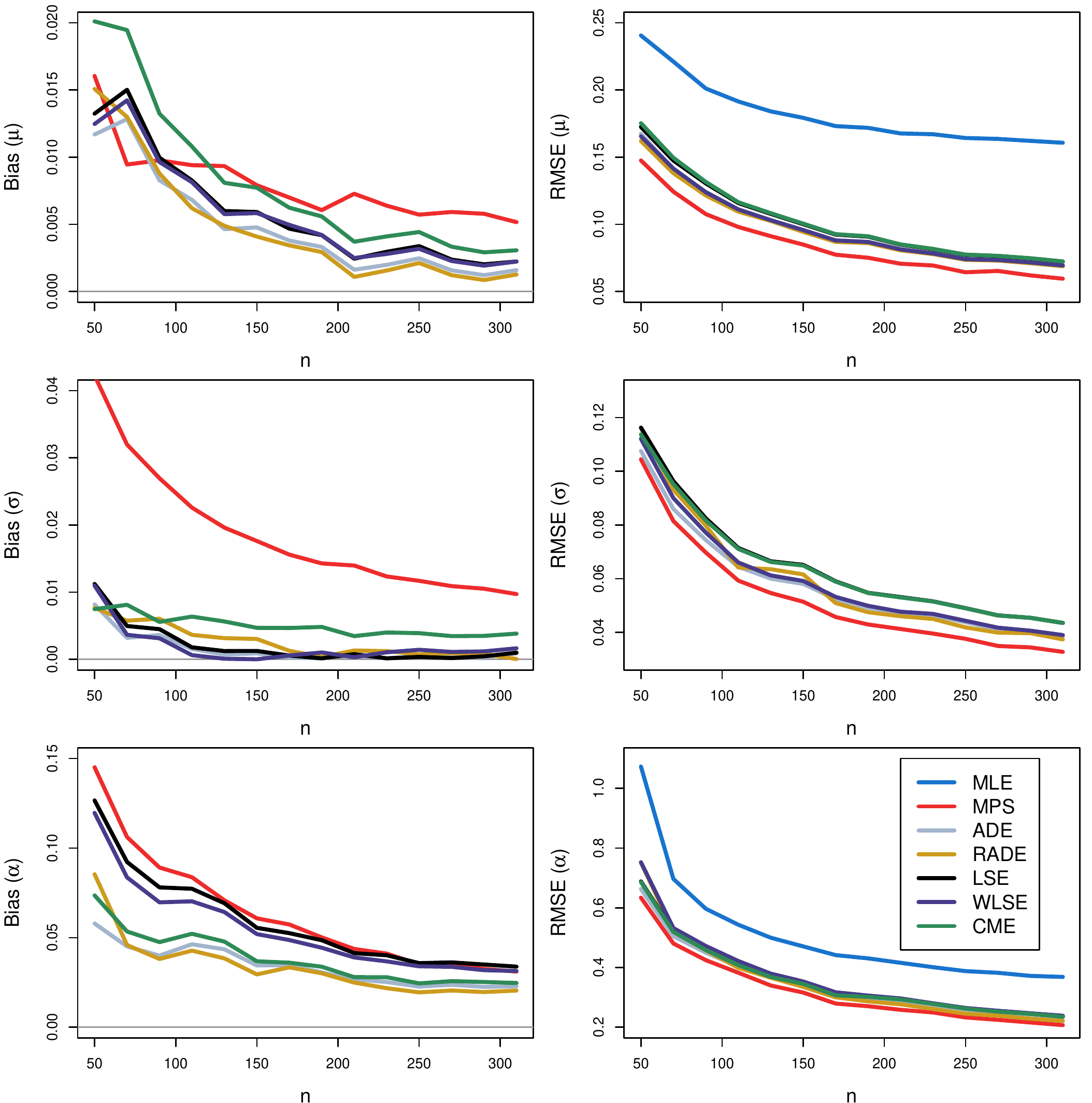}
\caption{Bias and MSE of the estimates of $\mu=0.5$, $\sigma=0.5$ and $\alpha=3$, for $N=10,000$ simulated samples of size $n$, using the following methods: 1-MLE, 2-MPS, 3-ADE, 4-RADE, 5-LSE, 6- WLSE, 7 - CME.} \label{fsimulation1}
\end{figure}

This simulated study's results shall return the most expected efficient estimation method conditioning on their estimations both Bias and MSE closer to zero. For this simulation study, we adopted the R software (R Core Team \citeyear{r2014}), and for the maximization method used package \textit{maxLik} and \textit{stats4} (Henningsen and Toomet, \citeyear{henningsen2011maxlik}). The chosen values of the simulation parameters were: $N = 10,000$ and $n = \{40, 60, 80,\cdots, 300 \}$. Due to lack of space, we will present the results only for $\{ \mu=0.5, \sigma=0.5, \alpha=3\}$ and $ \{ \mu=0, \sigma=1, \alpha=5\}$. Nonetheless, the following results are generalized by other choices of the vector of parameters $\boldsymbol{\theta}$. The estimation methods are considered under the same conditions in terms of samples, limit iterations numbers, and initial values. Here, we considered the true values as initial values. However, we provide a simple approach discussed in the next section to deal with real cases where good initial values are not available.

\begin{figure}[!h]
\centering
\includegraphics[scale=0.6]{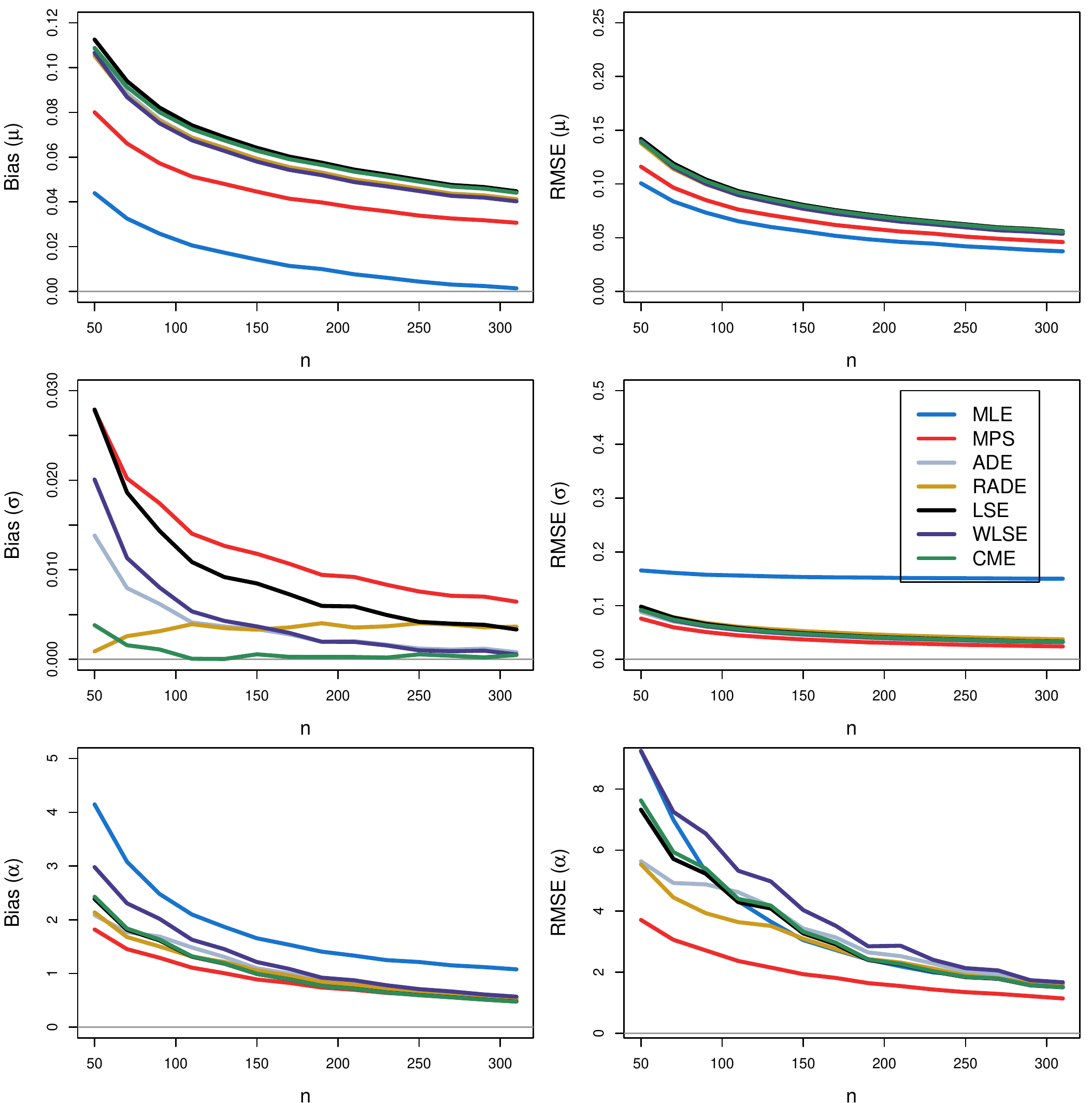}
\caption{Bias and MSE of the estimates of $\mu=0$, $\sigma=1$ and $\alpha=5$, for $N=10,000$ simulated samples of size $n$, using the following methods: 1-MLE, 2-MPS, 3-ADE, 4-RADE, 5-LSE, 6- WLSE, 7 - CME.} \label{fsimulation2}
\end{figure}

Figures \ref{fsimulation1} and \ref{fsimulation2}  present the performance of the estimators in terms of Bias and MSE for the parameters $\mu$, $\sigma$, and $\alpha$ using the MLEs, LSQs, WLQs, MPSs, CMEs, ADEs and RTADEs, with $N = 10.000$ simulated samples, and different values of $n$. It can be observed that the MLEs do not return adequate estimates for some parameter values and only converges for large samples of size. These results show a drawback in the current approach used to obtain the parameter estimates of ASN distribution. Although there is no uniform method that returns better estimates for all parameters and different parameter values, we observed that the ADEs obtained the best results in terms of minimum Bias and MSE. Additionally, obtaining an estimate for $\alpha$ is quite challenging to estimate given the influence from the parameter of location and scale $\mu$ and $\sigma$). In this approach, we faced fewer computational issues to obtain such estimates, and therefore we recommend using the ADEs to achieve the estimates for all practical purposes.

\section{Results}\label{results}

\begin{figure}[!ht]
    \centering
    \includegraphics[scale=0.5]{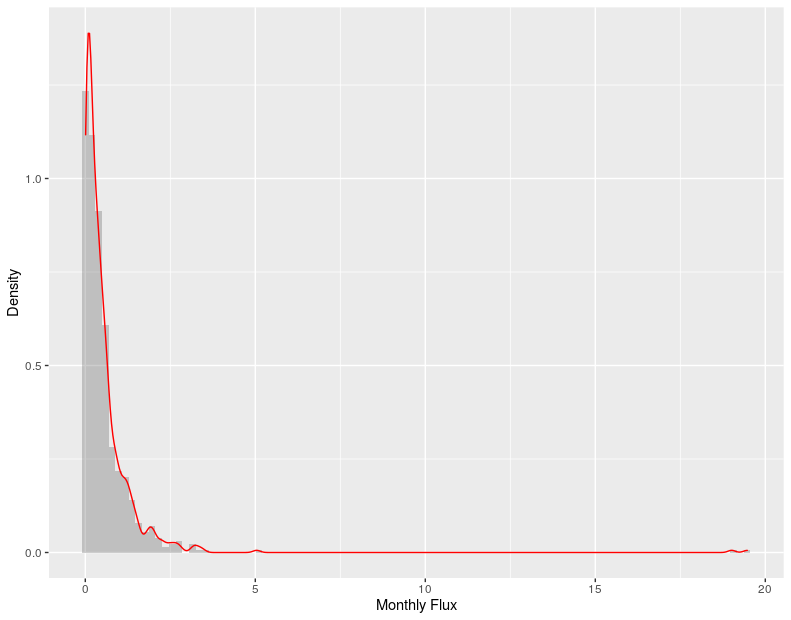}
    \caption{Empirical density function of the water flux from the 21 river/channels from the Copiapó neighborhood. The gray solid-shade represents the density (frequency) of each numerical records of the water flux, and the red solid-line a smooth adjusted function.}
    \label{fig:density}
\end{figure}

As we presented in Section \ref{data}, the motivation of this paper is driven by the Atacama Desert water flux. More precisely in the Copiapó neighborhood. Figure \ref{fig:density} shows the empirical density of this phenomenon, whereas a high concentration of low-values is presented (near to zero) although important events also captured in this 10-year time window, such as a big rain retrieving a large leptokurtosis.

Table \ref{tab:summary} presents the statistical summaries (minimum, 1st Quartile, Median, Mean, 3rd Quartile, maximum) of the water flux per month. Since the weather is very constant in the region, the seasonality across years is to be ignored since low flux is common (close values through the minimum of the months). Nonetheless, the cycle per month is essential given events like defrost at the end of spring/beginning of summer (higher values in the 3rd quartile in NOV and DEC), it is expected to receive more water in the system.

\begin{table}[htbp]
\centering
\caption{Summary Statistics of the Water Flux per month.}
\begin{tabular}{c|c|c|c|c|c|c|c}
\hline
 Month &    Min. &  1st Qu. &   Median &     Mean &  3rd Qu. &     Max. &     NA's \\ \hline
JAN & 0.02 & \textbf{0.06} & 0.31 & 0.5374 & 0.68 & 3.45 & 39 \\ 
FEB & 0.01 & \textbf{0.065} & 0.2 & 0.5165 & 0.6875 & 3.15 & 40 \\ 
MAR & 0.01 & \textbf{0.06} & 0.31 & 0.5449 & 0.85 & 3.24 & 37 \\ 
APR & 0.03 & 0.08 & 0.27 & 0.4494 & 0.5275 & 2.25 & 36 \\ 
MAY & 0.03 & 0.12 & 0.29 & \textbf{0.7859} & 0.55 & \textbf{19.47} & 47 \\ 
JUN & 0.02 & 0.12 & 0.35 & \textbf{0.9106} & 0.62 & \textbf{19.01} & 51 \\ 
JUL & 0.01 & 0.14 & \textbf{0.46} & 0.5636 & 0.64 & 2.58 & 53 \\ 
AUG & 0.01 & 0.1125 & 0.33 & 0.4692 & 0.6175 & 2.23 & 50 \\ 
SEP & 0.01 & 0.2025 & \textbf{0.45} & 0.5356 & 0.6775 & 2.66 & 52 \\ 
OCT & 0.02 & 0.1 & 0.37 & 0.5229 & \textbf{0.77} & 2.46 & 49 \\ 
NOV & 0.01 & 0.0775 & 0.365 & 0.5536 & \textbf{0.855} & 3.36 & 48 \\ 
DEC & 0.01 & 0.055 & 0.265 & 0.5639 & \textbf{0.7975} & 5.04 & 46 \\ \hline
\end{tabular}
\label{tab:summary}
\end{table}

\begin{figure}[!ht]
    \centering
    \includegraphics[width=\textwidth]{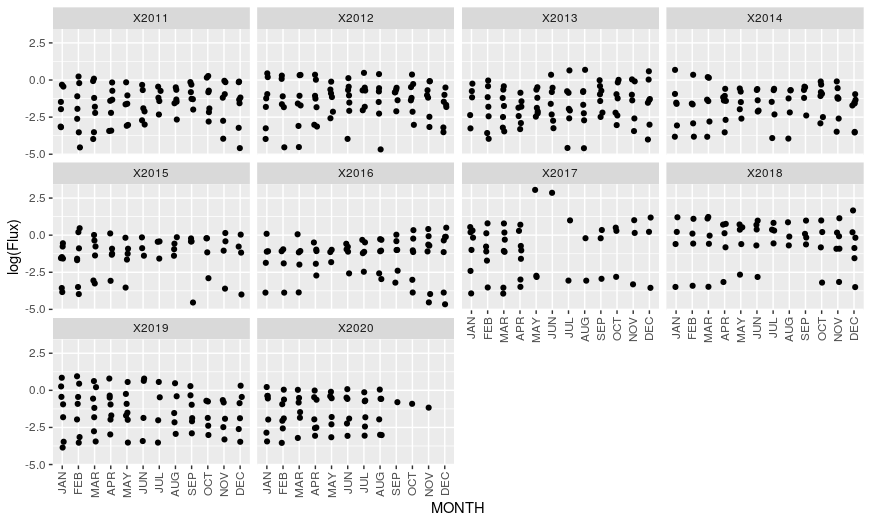}
    \caption{The logarithm of the water flux dispersion records (y-coordinates), per year (per panel), through the months (x-coordinates).}
    \label{fig:year}
\end{figure}

\begin{figure}
    \centering
    \includegraphics[width=\textwidth]{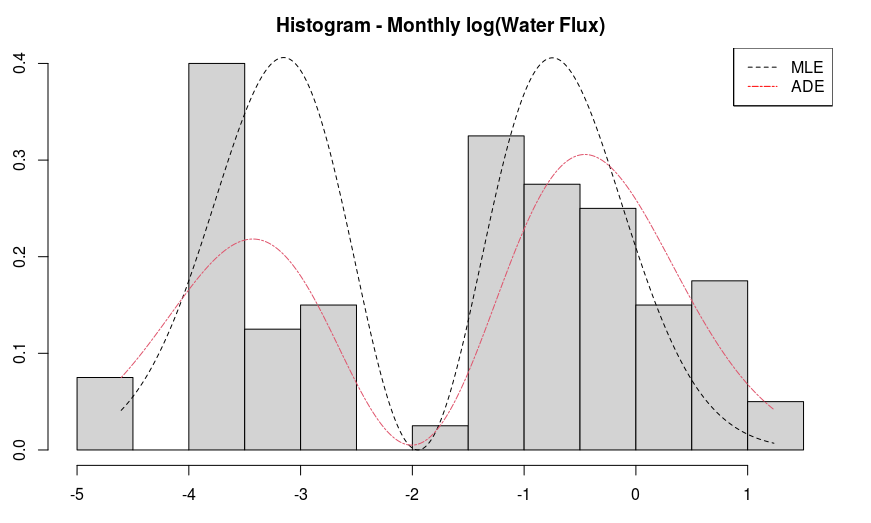}
    \caption{The empirical distribution of the log of the water flux, and its frequency represented by gray blocks. The black dashed line represents the adjusted ASN distribution based on the MLE ($\mu=-1.949,\sigma=0.85,\alpha=4147.07$), and the red dot-dashed line based on the ADE ($\mu=-1.879,\sigma=1.05,\alpha=-8.36$).}
    \label{fig:ASN-est}
\end{figure}

The use of the logarithm transformation has been used for a long time \cite{finney1941distribution}, though often obtained a normal distribution, some other situations, not the case, for instance, the dynamic of the water flux through the observed period in Figure \ref{fig:ASN-est}, showing the presence of bimodality in this data transformation, and its monthly representation by Figure \ref{fig:year}. It is essential to mention that the maximum historical values were in MAY and JUN (of 2017), related to the heavy rains that occurred in the region, also notable in the previous Table \ref{tab:summary}.

The empirical distribution of this phenomenon is visualized in Figure \ref{fig:ASN-est}, whereas the dashed-line represents the adjusted density functions, in black adopting the MLE and in red adopting the ADE. The initial values used to start the iteration procedures were obtained from
\begin{equation*}
\tilde{\mu}=\sum_{i=1}^{n}\frac{x_i}{n} \quad \mbox{and} \quad  \tilde{\sigma}=\sum_{i=1}^{n}\frac{(x_i-\tilde{\mu})^2}{n}.
\end{equation*}
while $\tilde{\alpha}$ is obtained from a grid search considering the range $(-10,-9.5,\ldots,9,10)$. The initial values of $\tilde{\mu}$ and $\tilde{\sigma}$ are clearly biased as they are the standar MLE for the normal distribution assuming that $\alpha=1$, on the other hand, the obtained values are not so far from the true value of the ASN, therefore, they can be used as good initial values only. Additionally, the Kolmogorov-Smirnov test of the MLE showed an statistic test $D = 0.094$, presenting a p-value of 0.5, whereas the ADE had a $D = 0.12$, with the associated p-value of 0.2 (suggesting the adequacy of the methods).

After confirming the ASN distribution's good-of-fitness, event occurrence can be associated with its density (or cumulative) probability function. For instance, extreme values are to be seen as Table \ref{tab:ASNprob} shows some exemplifications taking into account the 1\%, 10\%, 50\%, 99\%, and 99.99\%.

\begin{table}[!h]
\centering
\caption{Cumulative event probability based on the adjusted ASN distribution (using ADE).}
\begin{tabular}{c|c|c|c|c|c}
\hline
CUM Prob. & 1\% & 10\% & 50\% & 99\% & 99.99\% \\ \hline
Flux & 0.0059 & 0.0174 & 0.3396 & 1.5068 & 16.281 \\ \hline
\end{tabular}
\label{tab:ASNprob}
\end{table}

\section{Conclusion}\label{conclusion}
Uncertainty reveals a wide variety of processes and experiences, which may follow different rules, although different attributions of uncertainty such as external (disposition) versus internal (ignorance) are assessed by statistical inference, given philosophical interpretations of probability \cite{kahneman1982variants}. The utilities of each possible outcome lead to choosing rational actions regardless of the observed results' uncertainty.

The example brought by this paper is the water flux modeling related to an essential element stressed by the significant population growth and the increase in the demand for water supply (for agriculture, industrial process, mineral extraction, human consumption) \cite{w13060824}. Therefore, planning the logistic of these excessive water decisions, upon the probabilistic distribution, helps in unraveling this complex task \cite{jain2002short,Tu2021}, initiated by the analysis of the water flux, especially when environmental factors present limit sources towards the water level.

Thus, this work proposed the investigation towards comparing different inference methods towards the ASN probabilistic distribution, which shows a promising and flexible distribution. Bimodality was noticeable and skewed information observed in the historical series, nevertheless accommodated by the adopted probabilistic approach.

Big-data solution may now be implemented, using real-time analysis, once the process's probabilistic function was estimated and evidence towards the good-of-fitness was shown. Monitoring charts and other statistical process control (SPC) tools can also explore since parametric distribution is often adopted, and here shown elements onto the ASN distribution. Future works should expand into the reasoning towards quantile estimations associated with this problem with explainable features (in a regression structure), and forecasting may also be a further research motivation later.

\section*{Disclosure statement}

No potential conflict of interest was reported by the author(s)

\section*{Acknowledgments}

Diego Nascimento acknowledges the support from the S\~ao Paulo State Research Foundation (FAPESP process 2020/09174-5). Pedro L. Ramos acknowledge the support from the S\~ao Paulo State Research Foundation (FAPESP process 2017/25971-0). Francisco Louzada acknowledges support from the S\~ao Paulo State
Research Foundation (FAPESP Processes 2013/07375-0) and CNPq (grant no. 301976/2017-1).

\bibliographystyle{chicago}
\bibliography{reference}

\end{document}